\documentclass[prd,twocolumn,floatfix,amsmath,nofootinbib,amssymb,floatfix]{revtex4}
\usepackage{graphicx,color,dcolumn,booktabs,bm,multirow}
\usepackage{longtable,lscape}
\usepackage{txfonts}
\usepackage{overpic}
\usepackage{amssymb}
\usepackage{indentfirst}
\usepackage{feynmf}   
\usepackage{slashed}  
\usepackage{cases}
\usepackage{color}
\usepackage{multirow}
\usepackage{epstopdf}
\usepackage{graphicx,color,dcolumn,booktabs,bm}
\usepackage[colorlinks, citecolor=blue,anchorcolor=red,menucolor=red, linkcolor=red,filecolor=red,runcolor=red,urlcolor=blue,frenchlinks=red]{hyperref}

\begin{document}
\title{Determining the width of $D_{s0}^{*}(2317)$ by using $T_{c\bar{s}0}^{a}(2327)$ in a molecular frame}
\author{Zi-Li Yue$^{1,2}$}
\email{zili.yue@ge.infn.it}
\author{Quan-Yun Guo$^{1}$}
\author{Dian-Yong Chen$^{1,3}$\footnote{Corresponding author}} \email{chendy@seu.edu.cn}
\author{Elena Santopinto$^2$}
\email{elena.santopinto@ge.infn.it}
\affiliation{
 $^{1}$ School of Physics, Southeast University,  Nanjing 210094, China}
\affiliation{
 $^{2}$ INFN, Sezione di Genova, Via Dodecaneso 33, 16146 Genova, Italy}
\affiliation{$^3$Lanzhou Center for Theoretical Physics, Lanzhou University, Lanzhou 730000, P. R. China}
\begin{abstract}
Motivated by the recent observation of the open-charm tetraquark $T_{c\bar{s}0}^{a}(2327)$ by the LHCb Collaboration, as well as results from Lattice QCD calculations, we consider the $T_{c\bar{s}0}^{a}(2327)$ and the $D_{s0}^{*}(2317)$ as $DK$ molecular states, with $I(J^{P})$ equal to $1(0^{+})$ and $0(0^{+})$, respectively, and we investigate their strong decay behavior in an effective Lagrangian approach. Within the model parameter range, we can reproduce the $T_{c\bar{s}0}^{a}(2327)$ experimental decay width, with the assumption that the $D_{s}^{+}\pi^{0}$ is the dominant decay channel of the $T_{c\bar{s}0}^{a+}(2327)$. In the same parameter range, we can establish a stringent limitation for the decay width of the $D_{s0}^{*}(2317)$, which is $(63.0-209)~\mathrm{keV}$ being significantly smaller than the PDG upper limit value.

\end{abstract}

\maketitle

\section{Introduction}\label{sec:1}
Beyond the conventional $q\bar{q}$ mesons and $qqq$ baryons, the existence of more complex hadronic configurations, referred to as exotic states, was first proposed by Gell-Mann in 1964 \cite{Gell-Mann:2021nyh}. However, it was not until the 21st century, with the advancement of high-energy physics experimental techniques and the accumulation of experimental data, that candidates for exotic states began to be observed~\cite{BaBar:2003oey,CLEO:2003ggt,Belle:2003nnu,BaBar:2004cah,Belle:2011vlx,Belle:2004lle,Belle:2009and,LHCb:2020pxc,LHCb:2022lzp,LHCb:2021vvq,BESIII:2013ris,Belle:2013yex,BESIII:2020qkh,LHCb:2021uow,BESIII:2013ouc}. As one of the typical examples, the $D_{s0}^{*+}(2317)$ was first reported by the BABAR Collaboration in the inclusive $D_{s}^{+}\pi^{0}$ invariant mass distribution in the $e^{+}e^{-}$ annihilation process in 2003~\cite{BaBar:2003oey}. Later, the CLEO Collaboration confirmed the existence of the $D_{s0}^{*+}(2317)$ in the same process~\cite{CLEO:2003ggt}. The PDG reports an average mass of $m=2317.8\pm 0.5~\mathrm{MeV}$, an upper limit decay width of $3.8~\mathrm{MeV} $ and $I(J^{P})$ quantum numbers equal to $0(0^{+})$  ( $ J$  and $P$  need confirmation; in any case, $J^{P}$ is natural and the low mass is consistent with $0^{+}$,  as written by the PDG)~\cite{ParticleDataGroup:2024cfk}.  

The observed mass of the $D_{s0}^{*+}(2317)$ lies significantly below the prediction of the quark model~\cite{Godfrey:2003kg,Godfrey:1985xj}. In order to understand the low mass puzzle, various interpretations have been proposed, such as the $DK$ molecular state~\cite{Zhu:2019vnr,Navarra:2015iea,Albaladejo:2015kea,Guo:2014ppa,Xie:2010zza,Zhang:2006ix,Guo:2006fu,Barnes:2003dj,Chen:2004dy,Faessler:2007gv,Xiao:2016hoa,Cleven:2014oka,Albaladejo:2016hae,Fu:2021wde,Sun:2015uva,Guo:2008gp,Lutz:2007sk}, compact tetraquark state~\cite{Dmitrasinovic:2004cu,Dmitrasinovic:2005gc,Zhang:2006hv,Hayashigaki:2004st,Nielsen:2005zr} and $P$-wave charm-strange meson coupled to a $DK$ channel~\cite{Zhang:2024usz,Badalian:2007yr,Badalian:2008zz,Wu:2014era,Guo:2007up,Hwang:2004cd,Hwang:2005tm,Simonov:2004ar,Zhou:2011sp,Chen:2003jp}. Among these explanations, the $DK$ molecular scenario has attracted most attention, as the mass of the $D_{s0}^{*+}(2317)$ lies below the $DK$ threshold. Consequently, extensive investigations have been conducted in a $DK$ molecular scenario from various perspectives, including the production mechanisms~\cite{Zhu:2019vnr,Navarra:2015iea,Albaladejo:2015kea,Guo:2014ppa}, the mass spectrum~\cite{Xie:2010zza,Zhang:2006ix,Guo:2006fu,Barnes:2003dj,Chen:2004dy}, and the decay properties~\cite{Faessler:2007gv,Xiao:2016hoa,Cleven:2014oka,Albaladejo:2016hae,Fu:2021wde,Sun:2015uva,Guo:2008gp,Lutz:2007sk}. It is worth mentioning that in Ref.~\cite{Faessler:2007gv} the authors used an effective Lagrangian approach to study the strong and radiative decays of the $D_{s0}^{*+}(2317)$ in a $DK$ molecular frame, and the partial width of $D_{s0}^{*+} (2317) \to D_s \pi^0$ was estimated to be $(46.7- 111.9)$ keV on varying the scale parameter $\Lambda$ from 1 to 2 GeV.

Recently, an open-charm tetraquark candidate $T_{c\bar{s}}^{++}$ and its isospin partner $T_{c\bar{s}}^{0}$  have been observed by the LHCb Collaboration in the process  $D_{s1}(2460)^{+}\to D_{s}^{+}\pi^{+}\pi^{-}$ ~\cite{LHCb:2024iuo}, and their mass and width are reported to be,
\begin{eqnarray}
m&=&2327\pm13\pm13~\mathrm{MeV},\nonumber\\
\Gamma&=&96\pm16_{-23}^{+170}~\mathrm{MeV}.
\end{eqnarray}  
From the observed channels, the spin-parity and isospin of the $T_{c\bar{s}}^{++/0}$ are most likely $J^{P}=0^{+}$ and $I=1$~\cite{LHCb:2024iuo}. In accordance with the naming convention proposed by the LHCb Collaboration~\cite{Gershon:2022xnn}, we will refer to the triplet $T_{c\bar{s}0}^{a0}(2327)$, $T_{c\bar{s}0}^{a+}(2327)$ and $T_{c\bar{s}0}^{a++}(2327)$ as $T_{c\bar{s}0}^{a}(2327)$.

Like the $D_{s0}^{*+}(2317)$, the newly observed $T_{c\bar{s}0}^{a}(2327)$ state is also located near the $DK$ threshold, which makes the $DK$ molecular interpretations for the $D_{s0}^{*+}(2317)$ and the $T_{c\bar{s}0}^{a}(2327)$ even more intriguing. In Ref.~\cite{Gregory:2025ium}, Lattice QCD calculations support the identification of the $T_{c\bar{s}0}^{a}(2327)$ as a $DK$ molecular state. An amplitude analysis of the process $D_{s1}(2460)^{+}\to D_{s}^{+}\pi^{+}\pi^{-}$ based on final state interactions indicated that the $T_{c\bar{s}0}^{a}(2327)$ could be considered as a $DK$ molecular state~\cite{Wang:2024fsz}. 

It is interesting that, by assuming the $D_{s1}(2460)^{+}$ to be a $D^{*}K$ molecule, the authors of Ref.~\cite{Tang:2023yls} predicted a double-bump structure in the $\pi^{+}\pi^{-}$ invariant mass spectrum in the process $D_{s1}(2460)^{+}\to D_{s}^{+}\pi^{+}\pi^{-}$; this double-bump structure has just been observed by the LHCb Collaboration~\cite{LHCb:2024iuo}. In Ref.~\cite{Gregory:2025ium}, the authors studied the same process by also assuming the $D_{s1}(2460)^{+}$ to be a $D^{*}K$ molecule, and found a good agreement with the experimental data~\cite{LHCb:2024iuo}. According to the heavy quark spin symmetry ($D,D^{*}$), the $D_{s0}^{*+}(2317)$ has been also suggested ~\cite{Gregory:2025ium,Guo:2025vjw} to be a $DK$ molecule; for more discussions  see~\cite{Guo:2025vjw}.

Stimulated by the discovery of $T_{c\bar{s}0}^{a}(2327)$~\cite{LHCb:2024iuo} and by Lattice calculation~\cite{Gregory:2025ium}, in this study, we consider the $T_{c\bar{s}0}^{a+}(2327)$ as the isospin vector counterpart of the $D_{s0}^{\ast+}(2317)$ in a $DK$ molecular frame; we investigate the  $T_{c\bar{s}0}^{a+}(2327)$ and $D_{s0}^{*+}(2317)$ strong decay widths, by using an effective Lagrangian approach~\cite{Faessler:2007gv}. In addition, the measurement of the $T_{c\bar{s}0}^{a+}(2327)$ width can serve as a reliable foundation for determining the $D_{s0}^{*+}(2317)$ width in the same molecular frame. We first reproduce the $T_{c\bar{s}0}^{a+}(2327)$ experimental decay width, and find that, by using the same model parameter range, the width of the $D_{s0}^{*+}(2317)$ can be predicted.

This paper is organized as follows. The hadronic molecular structures of the $T_{c\bar{s}0}^{a+}(2327)$ and the $D_{s0}^{*+}(2317)$ are introduced in Section \ref{sec:2}. The strong decays of the $T_{c\bar{s}0}^{a+}(2327)$ and the $D_{s0}^{*+}(2317)$ are estimated in Section \ref{sec:3}. The numerical results and the discussions are presented in Section \ref{sec:4}, and the last section provides a short summary.

\section{Hadronic molecular structure}
\label{sec:2}

In the present study, the newly discovered $T_{c\bar{s}0}^{a}(2327)$ and the $D_{s0}^{*}(2317)$ (abbreviated to $T_{c\bar{s}0}^{a}$ and $D_{s0}^{*}$ hereafter) are considered as $DK$ molecular states with quantum numbers $I(J^{P})=1(0^{+})$ and $0(0^{+})$, respectively. The effective Lagrangians describing the interactions between $T_{c\bar{s}0}^{a+}(2327)/D_{s0}^{*+}(2317)$ and their components are, 
\begin{eqnarray}
\label{eq:mo}
\mathcal{L}_{T_{c\bar{s}0}^{a+}}(x)&=&g_{T_{c\bar{s}0}^{a+}} T_{c\bar{s}0}^{a+}(x) \int dy\Phi(y^2)\Big(D^{+}(x+\omega_{K}y)K^{0}(x-\omega_{D}y)\nonumber\\
&-&D^{0}(x+\omega_{K}y)K^{+}(x-\omega_{D}y)\Big)+\mathrm{h.c.},
\end{eqnarray}
\begin{eqnarray}
\label{eq:mo2}
\mathcal{L}_{D_{s0}^{*+}}(x)&=&g_{D_{s0}^{*+}}  D_{s0}^{*+}(x)\int dy\Phi(y^2)\Big(D^{+}(x+\omega_{K}y)K^{0}(x-\omega_{D}y)\nonumber\\
&+&D^{0}(x+\omega_{K}y)K^{+}(x-\omega_{D}y)\Big)+\mathrm{h.c.},
\end{eqnarray}
where $\omega_{i}=m_{i}/(m_{i}+m_{j})$ are the kinematical parameters, and $x$ and $y$ are the center of mass coordinate and the relative Jacobi coordinate, respectively. The correlation function $\Phi(y^2)$ characterizes the $DK$ molecules. The Fourier transformation of the $\Phi(y^2)$ is
\begin{equation}
\Phi(y^2)=\int \frac{d^4q}{(2\pi)^4}e^{-ipy}\tilde\Phi\left(-p^2\right).
\label{eq:fourier}
\end{equation}
The choice for $\Phi\left(-p^2\right)$ should satisfy the condition that describes the inner structure of the $DK$ molecular state and fall fast enough in the ultraviolet region of the Euclidean space. Here, we employ the correlation function in the Gaussian form~\cite{Faessler:2007gv,Faessler:2007us,Faessler:2008vc,Chen:2016byt,Dong:2017gaw,Gutsche:2010jf}, which is
\begin{equation}
\tilde\Phi(p_E^2)=\mathrm{exp}\left(-p_E^2/\Lambda^2\right),
\end{equation}
where $p_E$ represents the Euclidean Jacobi momentum and $\Lambda$ is a size parameter which
parametrizes the distribution of the $D$ and $K$ mesons inside the $DK$ molecules.

The coupling constants $g_{T_{c\bar{s}0}^{a+}}$ and $g_{D_{s0}^{*+}}$ in Eq.~(\ref{eq:mo}) can be determined by Weinberg's compositeness condition that $Z$ equals $0$~\cite{Weinberg:1962hj, Salam:1962ap,vanKolck:2022lqz},
\begin{eqnarray}
\label{eq:cc}
Z&=&1-\Pi^{\prime}(m^2)=0,
\end{eqnarray}
which implies that the $T_{c\bar{s}0}^{a+}(2327)$ and the $D_{s0}^{*+}(2317)$ are considered as pure molecular states. The $\Pi^{\prime}$ in Eq.~(\ref{eq:cc}) is the derivative of the mass operator of the $D_{s0}^{*+}(2317)$ or $T_{c\bar{s}0}^{a+}(2327)$.

\begin{figure}[t]
\begin{tabular}{c}
 \centering
  \includegraphics[width=8cm]{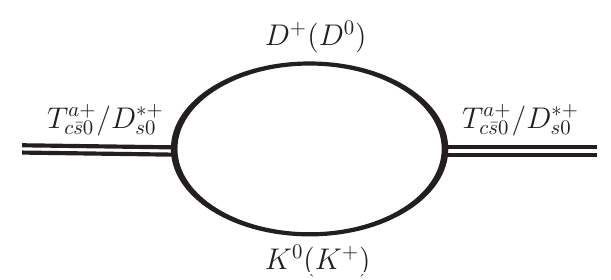} 
 \end{tabular}
\caption{The mass operators of the $T_{c\bar{s}0}^{a+}$ and $D_{s0}^{*+}$.\label{fig:mo}}
\end{figure}

According to the effective Lagrangians presented in Eq.~(\ref{eq:mo}) and Eq.~(\ref{eq:mo2}), the explicit form of the mass operators of the $T_{c\bar{s}0}^{a+}(2327)$ and the $D_{s0}^{*+}(2317)$ corresponding to Fig.~\ref{fig:mo} can be written as,
\begin{eqnarray}
\Pi(m_{T_{c\bar{s}0}^{a+}}^{2})&=&g_{T_{c\bar{s}0}^{a+}}^{2}\int\frac{d^{4}q}{(2\pi)^{4}}\tilde{\Phi}^{2}\left[-(q-\omega_{DK}P)^{2},\Lambda^{2}\right]\nonumber\\
&\times&\frac{1}{(p-q)^{2}-m_{K}^2}\frac{1}{q^{2}-m_{D}^{2}},\nonumber\\
\Pi(m_{D_{s0}^{*+}}^{2})&=&g_{D_{s0}^{*+}}^{2}\int\frac{d^{4}q}{(2\pi)^{4}}\tilde{\Phi}^{2}\left[-(q-\omega_{DK}P)^{2},\Lambda^{2}\right]\nonumber\\
&\times&\frac{1}{(p-q)^{2}-m_{K}^2}\frac{1}{q^{2}-m_{D}^{2}}.
\end{eqnarray}

\begin{figure}[t]
\begin{tabular}{cc}
  \centering
\includegraphics[width=4.2cm]{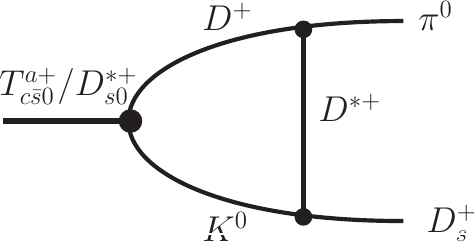}&
\includegraphics[width=4.2cm]{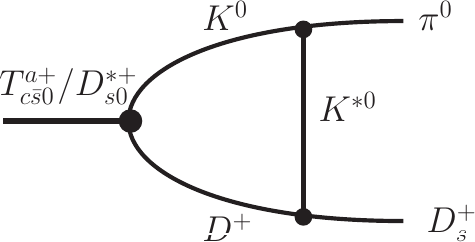}\\
 \\
 $(a)$ & $(b)$ \\ \\
\includegraphics[width=4.2cm]{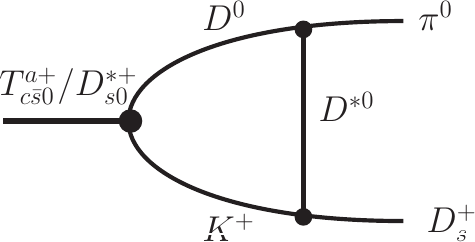}&
\includegraphics[width=4.2cm]{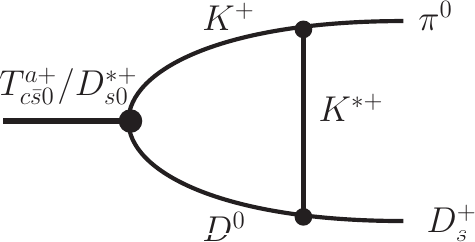}\\
 \\
 $(c)$ & $(d)$ 
\end{tabular}
\caption{The typical diagrams contributing to $T_{c\bar{s}0}^{a+}\to D_{s}^{+}\pi^{0}$ and $D_{s0}^{*+}\to D_{s}^{+}\pi^{0}$ at the hadron level.}
\label{fig:Decay}
\end{figure}

\begin{figure}[htb]
\begin{tabular}{cc}
  \centering
\includegraphics[width=4.2cm]{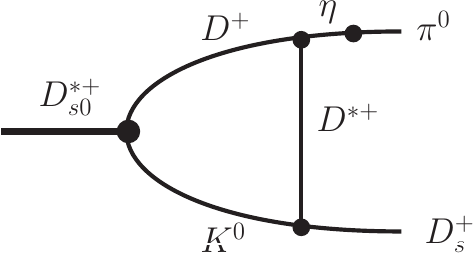}&
\includegraphics[width=4.2cm]{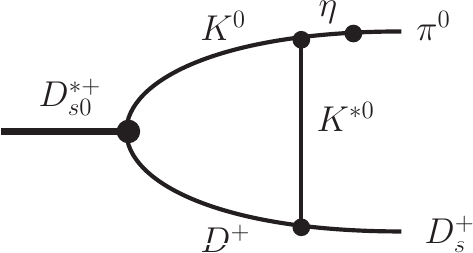}\\
 \\
 $(a)$ & $(b)$ \\ \\
\includegraphics[width=4.2cm]{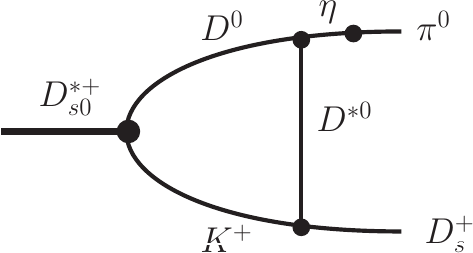}&
\includegraphics[width=4.2cm]{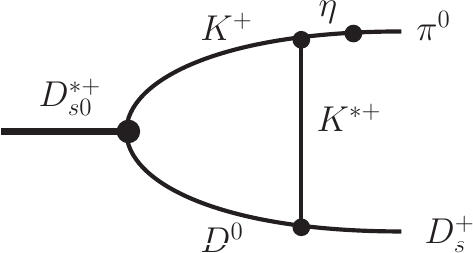}\\
 \\
 $(c)$ & $(d)$ 
\end{tabular}
\caption{The typical diagrams contributing to $D_{s0}^{*+}\to D_{s}^{+}\pi^{0}$ via $\eta$-$\pi^{0}$ mixing at the hadron level.}
\label{fig:mix}
\end{figure}

\section{Strong decays of the $T_{c\bar{s}0}^{a+}(2327)$ and the $D_{s0}^{*+}(2317)$}
\label{sec:3}

In the present study, we inverstigate the strong decay behaviors of the $T_{c\bar{s}0}^{a+}(2327)$ and the $D_{s0}^{*+}(2317)$ in the $DK$ molecular scenario. The possible strong decay processes are $T_{c\bar{s}0}^{a+}\to D_{s}^{+}\pi^{0}$ and $D_{s0}^{*+}\to D_{s}^{+}\pi^{0}$. The corresponding decay diagrams are collected in Fig.~\ref{fig:Decay} and Fig.~\ref{fig:mix}.

\subsection{Effective Lagrangian}
In the present paper, we use the effective Lagrangian approach introduced in Ref.~\cite{Faessler:2007gv} to estimate the diagrams in Fig.~\ref{fig:Decay} and Fig.~\ref{fig:mix}. Within the SU(4) symmetry, the Lagrangians of the $VPP$ interaction can be obtained by minimal substitution, and are~\cite{Faessler:2007gv,Haglin:2000ar,Haglin:1999xs,Lin:1999ad,Azevedo:2003qh,Kaymakcalan:1984bz,Gomm:1984at},
\begin{eqnarray}
\mathcal{L}_{D^{*}D_{s}K}&=&ig_{D^{*}D_{s}K}\Big(\bar{D}_{s}\partial_{\mu}K-K\partial_{\mu}\bar{D}_{s}\Big)D^{*\mu}+\mathrm{h.c.},\nonumber\\
\mathcal{L}_{DD_{s}K^{*}}&=&ig_{DD_{s}K^{*}}\Big(D\partial_{\mu}\bar{D}_{s}-\bar{D}_{s}\partial_{\mu}D\Big)K^{*\mu}+\mathrm{h.c.},\nonumber\\
\mathcal{L}_{D^{*}D\eta}&=&ig_{D^{*}D\eta}\Big(\eta\partial_{\mu}D-D\partial_{\mu}\eta\Big)\bar{D}^{*\mu}+\mathrm{h.c.},\nonumber\\
\mathcal{L}_{K^{*}K\eta}&=&ig_{K^{*}K\eta}\Big(K\partial_{\mu}\eta-\eta\partial_{\mu}K\Big)\bar{K}^{*\mu}+\mathrm{h.c.},\nonumber\\
\mathcal{L}_{K^{*}K\pi}&=&ig_{K^{*}K\pi}\Big(K\partial_{\mu}\vec{\tau}\cdot\vec{\pi}-\vec{\tau}\cdot\vec{\pi}\partial_{\mu}K\Big)\bar{K}^{*\mu}+\mathrm{h.c.},\nonumber\\
\mathcal{L}_{D^{*}D\pi}&=&ig_{D^{*}D\pi}\left(\vec{\tau}\cdot\vec{\pi}\partial_{\mu}D-D\partial_{\mu}\vec{\tau}\cdot\vec{\pi}\right)\bar{D}^{*\mu}+\mathrm{h.c.}.
\label{eq:lag}
\end{eqnarray}

For the process $D_{s0}^{*+}\to D_{s}^{+}\pi^{0}$, the contribution from $\eta-\pi^{0}$ mixing cannot be neglected. The interaction of the $\eta-\pi^{0}$ mixing term can be described by~\cite{Faessler:2007gv,Xiao:2016hoa},
\begin{eqnarray}
\mathcal{L}_{\eta\pi^{0}}=\mu\frac{m_{d}-m_{u}}{\sqrt{3}}\pi^{0}\eta,
\end{eqnarray}
where $m_u$ and $m_d$ are the $u$ and $d$ current quark masses, and $\mu$ is the condensate parameter. The mesons in Eq.~(\ref{eq:lag}) are from the 15-plets of the pseudoscalar mesons and vector mesons, which are
\begin{widetext}
\begin{eqnarray}
\mathcal{P}&=&
\begin{pmatrix}
\frac{\pi^{0}}{\sqrt{2}}+\frac{\eta}{\sqrt{6}}+\frac{\eta_{c}}{\sqrt{12}}&\pi^{+}&K^{+}&\bar{D}^{0}\\
\pi^{-}&-\frac{\pi^{0}}{\sqrt{2}}+\frac{\eta}{\sqrt{6}}+\frac{\eta_{c}}{\sqrt{12}}&K^{0}&D^{-}\\
K^{-}&\bar{K}^{0}&-\sqrt{\frac{2}{3}}\eta+\frac{\eta_{c}}{\sqrt{12}}&D_{s}^{-}\\
D^{0}&D^{+}&D_{s}^{+}&\frac{-3\eta_{c}}{\sqrt{12}}\nonumber\\
\end{pmatrix},\nonumber\\ \\
\mathcal{V}&=&
\begin{pmatrix}
\frac{\rho_{0}}{\sqrt{2}}+\frac{\omega^{\prime}}{\sqrt{6}}+\frac{J/\psi}{\sqrt{12}}&\rho^{+}&K^{*+}&\bar{D}^{*0}\\
\rho^{-}&-\frac{\rho_{0}}{\sqrt{2}}+\frac{\omega^{\prime}}{\sqrt{6}}+\frac{J/\psi}{\sqrt{12}}&K^{*0}&D^{*-}\\
K^{*-}&\bar{K}^{*0}&-\sqrt{\frac{2}{3}}\omega^{\prime}+\frac{J/\psi}{\sqrt{12}}&D_{s}^{*-}\\
D^{*0}&D^{*+}&D_{s}^{*+}&-\frac{3J/\psi}{\sqrt{12}}\nonumber
\end{pmatrix}.
\end{eqnarray}

\end{widetext}
\subsection{Decay amplitude}

With the effective Lagrangians listed above, we can obtain the amplitudes corresponding to the diagrams in Fig.~\ref{fig:Decay} (a)-(b), which are
\begin{eqnarray}
i\mathcal{M}_{a}&=&i^{3}\int\frac{d^{4}q}{(2\pi)^{4}}\Big[g_{T_{c\bar{s}0}^{a+}}\tilde{\Phi}(-p_{12}^{2},\Lambda^{2})\Big]\nonumber\\
&\times&\Big[ig_{D^{*}D\pi}\left(ip_{3}^{\mu}+ip_{1}^{\mu}\right)\Big]\Big[ig_{D^{*}D_{s}K}\left(-ip_{2}^{\nu}-ip_{4}^{\nu}\right)\Big]\nonumber\\
&\times&\frac{1}{p_{1}^{2}-m_{1}^{2}}\frac{1}{p_{2}^{2}-m_{2}^{2}}\frac{-g_{\mu\nu}+q_{\mu}q_{\nu}/m_{q}^{2}}{q^{2}-m_{q}^{2}},\nonumber
\end{eqnarray}
\begin{eqnarray}
i\mathcal{M}_{b}&=&i^{3}\int\frac{d^{4}q}{(2\pi)^{4}}\Big[g_{T_{c\bar{s}0}^{a+}}\tilde{\Phi}(-p_{12}^{2},\Lambda^{2})\Big]\nonumber\\
&\times&\Big[ig_{K^{*}K\pi}\left(-ip_{3}^{\mu}-ip_{1}^{\mu}\right)\Big]\Big[ig_{DD_{s}K^{*}}\left(ip_{2}^{\nu}+ip_{4}^{\nu}\right)\Big]\nonumber\\
&\times&\frac{1}{p_{1}^{2}-m_{1}^{2}}\frac{1}{p_{2}^{2}-m_{2}^{2}}\frac{-g_{\mu\nu}+q_{\mu}q_{\nu}/m_{q}^{2}}{q^{2}-m_{q}^{2}}.
\end{eqnarray}

It is easy to obtain the amplitudes corresponding to the (c) and (d) diagrams in Fig.~\ref{fig:Decay} by using the following replacements,
\begin{eqnarray}
\mathcal{M}_{c}&=&\mathcal{M}_{a}\Big|_{D^{+}\to D^{0},K^{0}\to K^{+},D^{*+}\to D^{*0}},\nonumber\\
\mathcal{M}_{d}&=&\mathcal{M}_{b}\Big|_{K^{0}\to K^{+},D^{+}\to D^{0},K^{*0}\to K^{*+}}.
\end{eqnarray}

The amplitudes contributing to the $D_{s0}^{*+}\to D_{s}^{+}\pi^{0}$  process via $\eta$-$\pi^{0}$ mixing, as shown in Fig.~\ref{fig:mix}, can be obtained by employing the following substitutions,
\begin{eqnarray}
\mathcal{M}_{a}^{mix}&=&\mathcal{M}_{a}\frac{m_{d}-m_{u}}{m_{s}-m}\frac{\sqrt{3}}{4}\Big|_{g_{D^{*}D\pi}\to g_{D^{*}D\eta}},\nonumber\\
\mathcal{M}_{b}^{mix}&=&\mathcal{M}_{b}\frac{m_{d}-m_{u}}{m_{s}-m}\frac{\sqrt{3}}{4}\Big|_{g_{K^{*}K\pi}\to g_{K^{*}K\eta}},\nonumber\\
\mathcal{M}_{c}^{mix}&=&\mathcal{M}_{c}\frac{m_{d}-m_{u}}{m_{s}-m}\frac{\sqrt{3}}{4}\Big|_{g_{D^{*}D\pi}\to g_{D^{*}D\eta}},\nonumber\\
\mathcal{M}_{d}^{mix}&=&\mathcal{M}_{d}\frac{m_{d}-m_{u}}{m_{s}-m}\frac{\sqrt{3}}{4}\Big|_{g_{K^{*}K\pi}\to g_{K^{*}K\eta}},
\end{eqnarray}
with $m=(m_{u}+m_{d})/2$. 

The amplitudes for the processes $T_{c\bar{s}0}^{a+}\to D_{s}^{+}\pi^{0}$ and $D_{s0}^{*+}\to D_{s}^{+}\pi^{0}$ are\footnote{It should be noted that the $\eta-\pi^0$ mixing diagrams in Fig.~\ref{fig:mix} also contribute to $T_{c\bar{s}0}^{a+}\to D_{s}^{+}\pi^{0}$, but their contributions are much smaller than those from the diagrams in Fig.\ref{fig:Decay}. Thus, the $\eta-\pi^0$ mixing contributions are neglected in the $T_{c\bar{s}0}^{a+}\to D_{s}^{+}\pi^{0}$ decay.}
\begin{eqnarray}
\label{eq:total}
\mathcal{M}_{T_{c\bar{s}0}^{a+}\to D_{s}^{+}\pi^{0}}&=&\mathcal{M}_{a}+\mathcal{M}_{b}-\mathcal{M}_{c}-\mathcal{M}_{d},\nonumber\\
\mathcal{M}_{D_{s0}^{*+}\to D_{s}^{+}\pi^{0}}&=&\mathcal{M}_{a}+\mathcal{M}_{b}+\mathcal{M}_{c}+\mathcal{M}_{d}\nonumber\\
&+&\mathcal{M}_{a}^{mix}+\mathcal{M}_{b}^{mix}+\mathcal{M}_{c}^{mix}+\mathcal{M}_{d}^{mix}.
\end{eqnarray}

With the amplitudes in Eq.~(\ref{eq:total}), the partial widths of $T_{c\bar{s}0}^{a+}\to D_{s}^{+}\pi^{0}$ and $D_{s0}^{*+}\to D_{s}^{+}\pi^{0}$ can be estimated by using the following expression,
\begin{eqnarray}
\Gamma_{T_{c\bar{s}0}^{a+}/D_{s0}^{*+}\to D_s^+ \pi^0}=\frac{1}{8\pi}\frac{|\vec{p}|}{m_{0}^{2}}\Big|\overline{\mathcal{M}_{T_{c\bar{s}0}^{a+}/D_{s0}^{*+}\to D_s^+ \pi^0}}\Big|^{2},
\end{eqnarray}
where the $m_{0}$ refers to the mass of the initial state, and $\vec{p}$ is the final-state momentum in the initial-state rest frame.

\section{Numerical Results and discussions}\label{sec3}
\label{sec:4}
\subsection{Coupling Constants}
The couplings $g_{D^{*}D\pi}=6.22$ and $g_{K^{*}K\pi}=3.12$ are taken from the measured decay widths of the processes $D^{*+}\to D^{+}\pi^{0}$ and $K^{*0}\to K^{0}\pi^{0}$~\cite{ParticleDataGroup:2024cfk}, respectively. According to the SU(4) symmetry, the $g_{D^{*}D\eta}$ and $g_{K^{*}K\eta}$ are
\begin{eqnarray}
g_{D^{*}D\eta}=\frac{f_{\pi}}{f_{\eta}\sqrt{3}}g_{D^{*}D\pi},~~~~g_{K^{*}K\pi}=\frac{f_{\pi}\sqrt{3}}{f_{\eta}}g_{K^{*}K\pi}.
\end{eqnarray}
The coupling $g_{D^{*}D_{s}K}=2.02$ is adopted from QCD sum rule calculations~\cite{Wang:2006ida,Bracco:2006xf}. Under SU(4) symmetry, we assume $g_{DD_{s}K^{*}}=g_{D^{*}D_{s}K}$ .

It is worth noting that, even though the SU(4) symmetry is strongly broken, our results remain meaningful, since the coupling constants are extracted from the experimental data. In this paper, the SU(4) framework is primarily employed to maintain consistency in the phase conventions of the mesons.
\begin{figure}[t]
  \centering
  \includegraphics[width=8.3 cm]{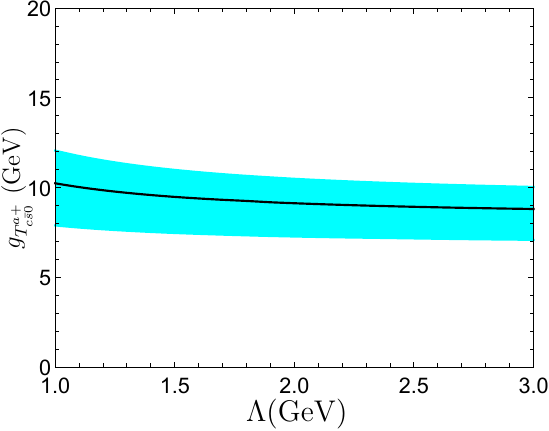}
\caption{(Color online.) The $g_{T_{c\bar{s}0}^{a+}}$ coupling constant, which depends on the $\Lambda$ model parameter, takes into account the uncertainty in the experimental mass~\cite{LHCb:2024iuo}.}\label{fig:coupling2327}
\end{figure}  

\begin{figure}[t]
  \centering
  \includegraphics[width=8.3 cm]{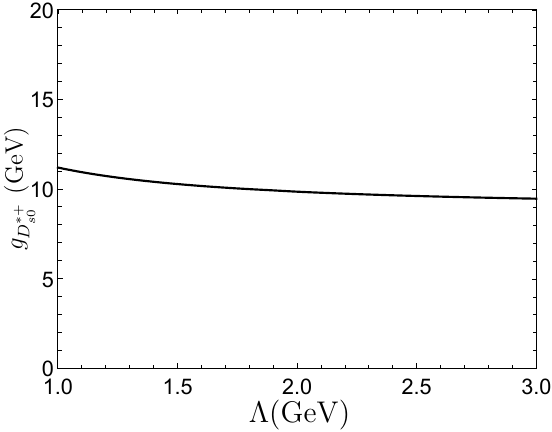}
\caption{The $g_{D_{s0}^{*+}}$ coupling constant as a function of the $\Lambda$ model parameter.}\label{fig:coupling2317}
\end{figure}
The coupling constants between the molecular states $T_{c\bar{s}0}^{a+}(2327)$ and $D_{s0}^{*+}(2317)$ and their components can be determined by employing the compositeness condition in Eq.~(\ref{eq:cc}). The empirical value of the model parameter $\Lambda$ is typically in the order of $1~\mathrm{GeV}$~\cite{Faessler:2007gv,Faessler:2007us,Faessler:2008vc}. In the present paper, we vary $\Lambda$ within the range from $1.0~\mathrm{GeV}$ to $3.0~\mathrm{GeV}$. The $g_{T_{c\bar{s}0}^{a+}}$ and $g_{D_{s0}^{*+}}$ coupling constants, which depend on $\Lambda$ parameter, are presented in Fig.~\ref{fig:coupling2327} and Fig.~\ref{fig:coupling2317}, respectively. The $g_{T_{c\bar{s}0}^{a+}}$ coupling constant is estimated by taking into account the experimental uncertainty in the mass (a lower mass corresponds to a larger coupling constant). In Fig.~\ref{fig:coupling2327}, inside the $\Lambda$ parameter range considered, the $g_{T_{c\bar{s}0}^{a+}}$ decreases from $12.06~\mathrm{GeV}$ to $10.04~\mathrm{GeV}$ (on using the lower limit of the mass), from $10.24~\mathrm{GeV}$ to $8.81~\mathrm{GeV}$ (on using the central value of the mass), and from $7.89~\mathrm{GeV}$ to $7.09~\mathrm{GeV}$ (on using the upper limit of the mass). In Fig.~\ref{fig:coupling2317}, the $g_{D_{s0}^{*+}}$ coupling constant decreases from $11.19~\mathrm{GeV}$ to $9.46~\mathrm{GeV}$ when the $\Lambda$ parameter increases from $1.0~\mathrm{GeV}$ to $3.0~\mathrm{GeV}$; in this case, the uncertainty of the $D_{s0}^{\ast +}(2317)$ mass is not taken into consideration, since it is very small. 


\subsection{Decay widths}
\begin{figure}[t]
  \centering
  \includegraphics[width=8.25 cm]{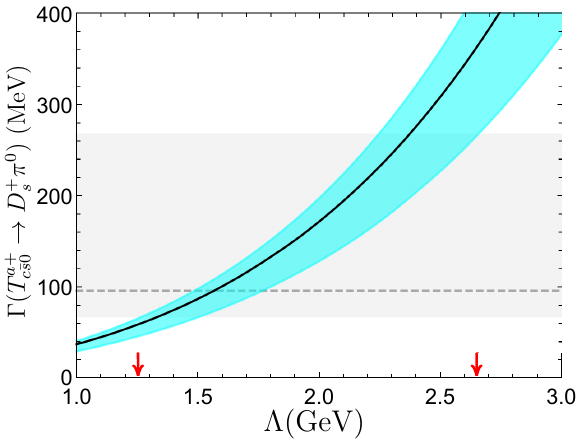}
\caption{(Color online.) The partial width of the $T_{c\bar{s}0}^{a+}\to D_{s}^{+}\pi^{0}$ decay. The solid black curve is the estimation using the central value of the $T_{c\bar{s}0}^{a+}(2327)$ mass, while the cyan band are the estimations including the mass uncertainty of $T_{c\bar{s}0}^{a+}(2327)$. The gray dashed line and the light gray band represent the experimental width of the $T_{c\bar{s}0}^{a+}(2327)$ measured by the LHCb Collaboration.}\label{fig:width2327}
\end{figure}

\begin{figure}[t]
  \centering
  \includegraphics[width=8.3 cm]{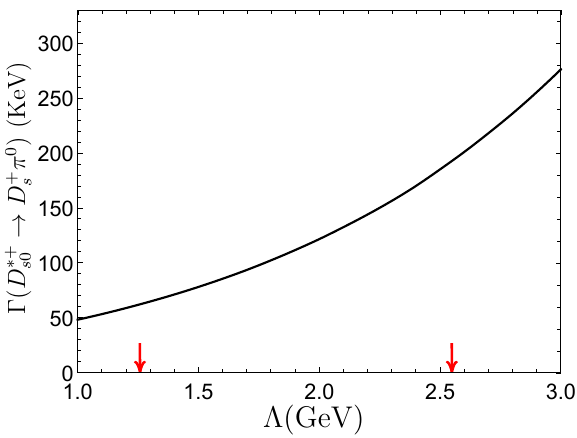}
\caption{The partial decay width of the isospin violation process $D_{s0}^{*+}\to D_{s}^{+}\pi^{0}$ depending on the parameter $\Lambda$ considering the $\eta$-$\pi^{0}$ mixing.}\label{fig:width2317}
\end{figure}

Fig.~\ref{fig:width2327} shows the partial decay widths of the process $T_{c\bar{s}0}^{a+}\to D_{s}^{+}\pi^{0}$ according to the $\Lambda$ parameter and the mass of the $T_{c\bar{s}0}^{a+}(2327)$.  The gray dashed line and the light gray band represent the experimental width of the $T_{c\bar{s}0}^{a+}(2327)$ measured by the LHCb Collaboration, which is $96\pm16_{-23}^{+170}~\mathrm{MeV}$. Within the assumption that the $T_{c\bar{s}0}^{a+}(2327)$ dominantly decays into $D_{s}^{+}\pi^{0}$, we can reproduce the $T_{c\bar{s}0}^{a+}(2327)$ width in the parameter range considered. Specifically, the $\Lambda$ parameter range is $1.34-2.37~\mathrm{GeV}$, $1.51-2.65~\mathrm{GeV}$ and $1.27-2.24~\mathrm{GeV}$ when the mass of the $T_{c\bar{s}0}^{a+}(2327)$ is $2327~\mathrm{MeV}$, $2345.38~\mathrm{MeV}$ and $2308.62~\mathrm{MeV}$, respectively. Considering both the uncertainties in the mass and the width of the $T_{c\bar{s}0}^{a+}(2327)$, the model parameter $\Lambda$ is determined to be 1.27-2.65 GeV, as indicated by the red arrows in Fig.~\ref{fig:width2327}.


Fig.~\ref{fig:width2317} presents the partial width of the $D_{s0}^{*+}\to D_{s}^{+}\pi^{0}$ process, according to the parameter $\Lambda$. In the estimation of the isospin violation process, the contribution of the $\eta$-$\pi^{0}$ mixing is also included, as discussed in Ref.~\cite{Faessler:2007gv}. The width of $D_{s0}^{*+}\to D_{s}^{+}\pi^{0}$ increases from $48.3~\mathrm{KeV}$ to $276.26~\mathrm{KeV}$ when  $\Lambda$ varies from $1.0~\mathrm{GeV}$ to $3.0~\mathrm{GeV}$. We observe that $\Gamma(D_{s0}^{*+}\to D_{s}^{+}\pi^{0})=48.3-121.7~\mathrm{KeV}$ when $\Lambda$ increses from $1.0~\mathrm{GeV}$ to $2.0~\mathrm{GeV}$; the slight discrepancy between our estimation and the results in Ref.~\cite{Faessler:2007gv} only arises from the slight differences in the values of the coupling constants.

In the $DK$ molecular scenario, the significant difference between the decay widths of the $T_{c\bar{s}0}^{a+}(2327)$ and the $D_{s0}^{*+}(2317)$ can be explained simultaneously. Furthermore, in the present paper, we also try to provide limitations on the width of the $D_{s0}^{*+}(2317)$ by using the parameter range determined by the decay width of the $T_{c\bar{s}0}^{a+}(2327)$. In the determined parameter range $1.27-2.65~\mathrm{GeV}$, we have the upper and lower limits of the width of the $D_{s0}^{\ast+}(2317)$ at $209~\mathrm{KeV}$ and  $63.0~\mathrm{KeV}$, respectively. To summarize, in the $DK$ molecular frame, we obtain a narrower width range for the $D_{s0}^{*+}(2317)$ than the PDG value $<3.8~\mathrm{MeV}$~\cite{ParticleDataGroup:2024cfk}. More precise measurements of the $T_{c\bar{s}0}^{a+}(2327)$ resonance would allow a more accurate determination of the model parameter. Subsequently, this parameter could be used to estimate the width of the $D_{s0}^{\ast +}(2317)$. Comparison of the estimated width of the $D_{s0}^{\ast +}(2317)$ obtained through further precise measurements can serve as a crucial test of the $DK$ molecular interpretations of the $T_{c\bar{s}0}^{a+}(2327)$ and the $D_{s0}^{\ast +}(2317)$.

\section{Summary}\label{sec4}
Stimulated by the observation of the open-charm tetraquark $T_{c\bar{s}0}^{a}(2327)$ by the LHCb Collaboration and by Lattice calculations~\cite{LHCb:2024iuo,Gregory:2025ium}, we investigate the possibility that the $T_{c\bar{s}0}^{a+}(2327)$ is a $DK$ molecular state from the perspective of strong decays by using an effective Lagrangian approach. In the present study, we consider the $T_{c\bar{s}0}^{a+}(2327)$ as the isospin vector counterpart of the $D_{s0}^{\ast +}(2317)$, and estimate the partial widths of the $T_{c\bar{s}0}^{a+}\to D_{s}^{+}\pi^{0}$ and $D_{s0}^{*+}\to D_{s}^{+}\pi^{0}$ decays in the $DK$ molecular scenario. Within the parameter range considered, we can reproduce the decay width of the $T_{c\bar{s}0}^{a+}(2327)$, and by using the same parameter range, we establish a stringent limitation to the decay width of the $D_{s0}^{*+}(2317)$, finding it to be between  $63.0$ and $209~\mathrm{keV}$, which is significantly smaller than the upper limit value indicated in the PDG.

\section{ACKNOWLEDGMENTS} 
This study is partly supported by the National Natural Science Foundation of China under Grant Nos. 12175037 and 12335001, and is supported, in part, by National Key Research and Development Program under contract No. 2024YFA1610503. Zi-Li Yue is also supported by the SEU Innovation Capability Enhancement Plan for Doctoral Students (Grant No. CXJH$\_$SEU 24135) and the China Scholarship Council (Grant No. 202406090305).

\end{document}